\documentclass[aps,prl,reprint,groupedaddress,superscriptaddress,amsfonts,amssymb,amsmath,showpacs,floatfix]{revtex4-1}
\usepackage{color}
\usepackage{enumerate}
\usepackage{graphicx}
\usepackage[rightcaption]{sidecap}
\usepackage{times}
\usepackage{float}

\newcommand{\BEATBOX}{\texttt{BeatBox}}   
\newcommand{\DXSPIRAL}{\texttt{DXSpiral}} 
\newcommand{\eg}[1]
  {{\it e.g.\/}\ifx#1.\else\expandafter#1\fi}
\newcommand{\eq}[1]{(\ref{eq:#1})}
\newcommand{\eqlabel}[1]{\label{eq:#1}}
\def\eqreftwo(#1,#2){(\ref{eq:#1},\ref{eq:#2})}
\newcommand{\eqtwo}[1]{\eqreftwo(#1)}
\newcommand{\etal}[1]
  {{\it et al.\/}\ifx#1.\else\expandafter#1\fi}
\newcommand{\Fig}[1]{Fig.~\figref{#1}}            
\newcommand{\fig}[1]{Fig.~\figref{#1}}            
\newcommand{\figref}[1]{\ref{fig:#1}}             
\newcommand{\figlabel}[1]{\label{fig:#1}}         
\newcommand{\ie}[1]
  {{\it i.e.\/}\ifx#1.\else\expandafter#1\fi}
\newcommand{\tab}[1]{Table~\ref{tab:#1}}          
\newcommand{\tablabel}[1]{\label{tab:#1}}         

\newcommand{\dblfigure}[3]
  {\begin{figure*}[tbp]\includegraphics{#1}\caption[]{#2}\figlabel{#3}\end{figure*}}
\newcommand{\sglfigure}[3]
  {\begin{figure}[tbp]\includegraphics{#1}\caption[]{#2}\figlabel{#3}\end{figure}}
\newcommand{\sidefigure}[3]%
  {\begin{SCfigure*}\includegraphics{#1}\caption[]{#2}\figlabel{#3}\end{SCfigure*}}

\renewcommand{\@}{\partial}             
  \newcommand{\<}{\langle}              
\renewcommand{\>}{\rangle}              
\newcommand{\const}{\mathrm{const}}     
\newcommand{\Df}[2]{\frac{\d{#1}}{\d{#2}}} 
\renewcommand{\d}{\mathrm{d}}            
\newcommand{\df}[2]{\frac{\@{#1}}{\@{#2}}} 
\newcommand{\diag}{\mathrm{diag}}       
\DeclareMathOperator{\dirac}{\delta}    
\newcommand{\E}[1]{\times10^{#1}}       
\newcommand{\e}{{\notcolor\mathrm{e}}} 
\DeclareMathOperator{\Heav}{\Theta}     
\renewcommand{\i}{{\notcolor\mathrm{i}}}           
\newcommand{\inner}[2]
  {\left\<#1\, , \,#2\right\>}
\newcommand{\Mx}[1]{
\left[\begin{array}{cccccccc}#1\end{array}\right]}
\newcommand{\mx}[1]{\mathbf{#1}}        
\renewcommand{\O}[1]{\mathcal{O}\left(#1\right)}  
\newcommand{\Real}{\mathbb{R}}          
\newcommand{\Tr}{^{\mathrm{T}}}          
\newcommand{\Z}{\mathbb{Z}}             

\newcommand{\notcolor}{}
\newcommand{\+}[2]{\def#1{{\notcolor#2}}}
\newcommand{\1}[2]{\def#1##1{{\notcolor#2}}}

\+{\angled}{\vartheta_0}        
\+{\balp}{\boldsymbol{\alpha}}  
\+{\bzoeps}{p}                  
\+{\bzof}{f}                    
\+{\bzoq}{q}                    
\+{\cma}{\vartheta}             
\+{\cmr}{\rho}                  
\+{\cocl}{\kappa}               
\+{\cra}{\theta}                
\+{\crr}{r}                     
\+{\D}{\mx{D}}                  
\+{\dists}{d}                   
\+{\distsc}{\dists^*}           
\+{\distd}{\ell}                
\+{\distdc}{\ell^*}             
\+{\eps}{\epsilon}              
\+{\F}{F}                       
\+{\Fun}{F}                     
\+{\Fx}{F_x}                    
\+{\Fy}{F_y}                    
\+{\Fs}{S}                      
\+{\Fsx}{S_x}                   
\+{\Fsy}{S_y}                   
\+{\Fdr}{F_r}                   
\+{\Fda}{F_a}                   
\+{\Fdac}{F_a^*}                
\+{\Ft}{T}                      
\+{\Ftx}{T_x}                   
\+{\Fty}{T_y}                   
\+{\f}{\mx{f}}                  
\+{\fhnalp}{\alpha}             
\+{\fhnbet}{\beta}              
\+{\fhngam}{\gamma}             
\+{\fun}{f}                     
\+{\g}{g}                       
\+{\gg}{h}                      
\+{\H}{H}                       
\+{\Ht}{\tilde{H}}              
\+{\Hin}{H_i}                   
\+{\Hout}{H_o}                  
\+{\h}{\mx{h}}                  
\+{\K}{K}                       
\+{\L}{\mathcal{L}}             
\+{\phase}{\phi}                
\+{\RFS}{\mx{W}}                
\+{\Ri}{R_{\textrm{d}}}         
\+{\r}{\vec r}                  
\+{\t}{t}                       
\+{\U}{\mx{U}}                  
\+{\u}{\mx{u}}                  
\+{\v}{\mx{v}}                  
\+{\vF}{{\vec{\F}}}             
\+{\vR}{{\vec R}}               
\+{\veps}{\mu}                  
\+{\vr}{{\vec r}}               
\+{\wedge}{\psi}                
\+{\width}{w}                   
\+{\widthc}{\width^*}           
\+{\X}{X}                       
\+{\Y}{Y}                       
\+{\x}{x}                       
\+{\y}{y}                       
\+{\z}{z}                       
\+{\zmin}{z_{\min}}             
\+{\zmax}{z_{\max}}             
\+{\zz}{z_*}                    
\+{\xs}{x_s}                    
\+{\xleft}{x_{\ell}}            
\+{\xright}{x_r}                
\+{\xd}{x_\textrm{d}}           
\+{\yd}{y_\textrm{d}}           

\+{\ii}{i'}                     
\+{\ix}{i}                      
\+{\ji}{j'}                     
\+{\jx}{j}                      
\+{\ki}{k'}                     
\+{\kx}{k}                      
\+{\N}{N}                       
\+{\Nr}{N_{\rho}}               
\+{\Nt}{N_{\theta}}             
\+{\Nx}{N_x}                    
\+{\Ny}{N_y}                    
\+{\Nz}{N_z}                    
\+{\Nzone}{N_{z,1}}             
\+{\Nztwo}{N_{z,2}}             
\+{\rp}{\rho_{\max}}            

\begin{document}
\title{Drift of scroll waves in thin layers caused by thickness
  features: \\ asymptotic theory and numerical simulations}

\author{I.V.~Biktasheva}
\affiliation{Department of Computer Science, University of Liverpool, 
     Liverpool L69 3BX, UK}

\author{H.~Dierckx}
\affiliation{Department of Mathematical Physics and Astronomy, 
  Ghent University, 
  9000 Ghent, Belgium}
     
\author{V.N.~Biktashev}
\affiliation{College of Engineering, Mathematics and Physical Sciences, University of Exeter, Exeter EX4 4QF, UK}

\date{\today}

\begin{abstract}
  A scroll wave in a very thin layer of excitable medium is similar to a
  spiral wave, but its behaviour is affected by the layer geometry. We
  identify the effect of sharp variations of the layer thickness,
  which is separate from filament tension and curvature-induced drifts
  described earlier. We outline a two-step asymptotic theory describing
  this effect, including asymptotics in the layer thickness and
  calculation of the drift of so perturbed spiral waves using response functions. As
  specific examples, we consider drift of scrolls along thickness
  steps, ridges, ditches, and disk-shaped thickness
  variations. Asymptotic predictions agree with numerical
  simulations. 
\end{abstract}

\pacs{%
  02.70.-c, 
  05.10.-a, 
  82.40.Bj,
  82.40.Ck, 
  87.10.-e 
}

\maketitle


Spiral waves in two dimensions (2D) and scroll waves in
  three dimensions (3D) are regimes of self-organization observed in
  physical, chemical and biological spatially-extended dissipative
  systems with excitable or self-oscillatory properties \cite {%
    Zhabotinsky-Zaikin-1971,
    *Allessie-etal-1973,
    *Alcantara-Monk-1974,
    *Gorelova-Bures-1983,
    *Madore-Freedman-1987,
    *Jakubith-etal-1990,
    *Lechleiter-etal-1991,
    *Frisch-etal-1994,
    *Cross-Hohenberg-1993%
  }. A particularly important example are the re-entrant waves of
  excitation underlying arrhythmias in the
  heart~\cite{Fenton-etal-2008}. In nature, 2D systems often are very
  thin 3D layers of the medium, so the dynamic fields vary only
  slightly in the transmural direction.  The geometry of a layer affects
  the dynamics of scroll waves via the well known phenomena of scroll wave
  filament tension~\cite{Biktashev-etal-1994} and surface curvature of
  the layer~\cite{Dierckx-etal-2013}, which cause scroll waves to
  drift to or from thinner regions and more curved regions
  respectively. There are, however, effects not reducible to these
  phenomena, and rather related to \emph{sharp features} of the layer
  thickness.  \Fig{step3d} shows a paradoxical example of a scroll
  wave with a \emph{positive} filament tension first attracted towards
  the \emph{thicker} part of the layer and then drifting along the
  thickness step.  There is experimental evidence that sharp thickness
  variations can play a significant role in atrial
  fibrillation~\cite{Wu-etal-1998-CR,Yamazaki-etal-2012-CVR}. %

In this Letter, we present an asymptotic theory of drift
  of scroll waves caused by variations of layer thickness. Predictions
  of this theory are quantitatively confirmed by direct numerical
  simulations for two selected archetypical models, one excitable and
  one self-oscillatory. We demonstrate that sharp variations can
  produce drifts that are not reducible to filament tension and
  surface curvature. The details of these drifts depend on the
  reaction-diffusion kinetics, as well as the size, geometry and
  position of the thickness feature. A typical motif, observed for
  both selected models, is that a scroll is first attracted towards a
  sharp thickness variation and then drifts along or around it.%


We start from a generic homogeneous isotropic reaction-diffusion
system in 3D:
\begin{align} \eqlabel{3D-pde}
  & \v_\t = \f(\v) + \D \nabla^2\v,
\end{align}
where $\v=[u(\r,\t), v(\r,\t)]^T$, $\r=(\x,\y,\z)$. 
In numerical examples, we use 
excitable FitzHugh-Nagumo system~\cite{%
  FitzHugh-1961,%
  *Nagumo-etal-1962,%
  *Winfree-1991-C%
}, with kinetics
\begin{align}
  \f: \Mx{u\\v} \mapsto \Mx{
    \fhnalp^{-1}(u-u^3/3-v) \\
    \fhnalp \, (  u + \fhnbet - \fhngam v )
  }\eqlabel{FHN}
\end{align}
for $\fhnalp=0.3$, $\fhnbet=0.68$, $\fhngam=0.5$, and 
$\D=\diag(1,0)$, and self-oscillatory Oregonator model of the
  Belousov-Zhabotinsky reaction \cite{%
  Field-Noyes-1974,%
  *Tyson-Fife-1980%
}, with kinetics
\begin{align}
  \f: \Mx{u\\v} \mapsto \Mx{
    \bzoeps^{-1}\left( u(1-u) - \bzof v
      \frac{u-\bzoq}{u+\bzoq}\right) \\
    u - v 
  } \eqlabel{BZO}
\end{align}
for $\bzoeps=0.1$, $\bzof=1.5$, $\bzoq=0.002$, and
$\D=\diag(1,0.6)$~\cite{epaps}.

\sglfigure{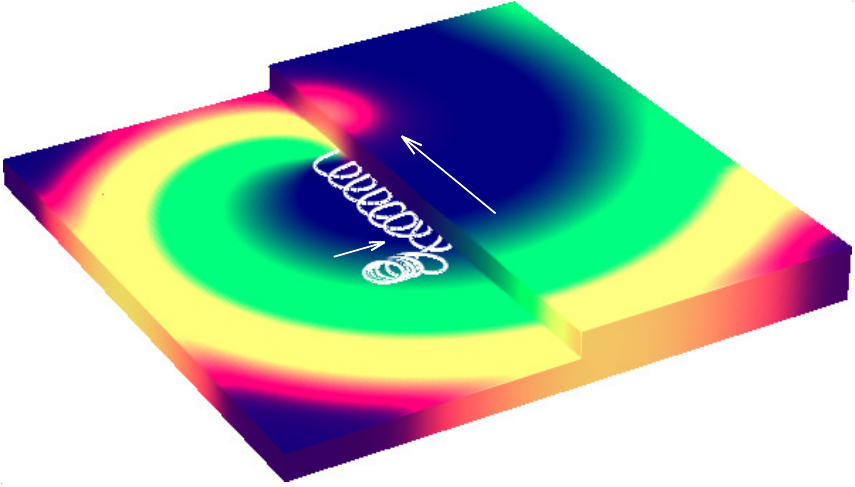}{%
  Surface view of a scroll wave in a thin layer of excitable medium
  described by FitzHugh-Nagumo system~\eqtwo{3D-pde,FHN}, with a stepwise
  variation of thickness. White curve is the trace of the vortex filament
  appearing at the top surface~\cite{epaps}. %
}{step3d}
 
We consider system~\eq{3D-pde} in a thin layer,
$\z\in[\zmin(\x,\y),\zmax(\x,\y)]$, $(\x,\y) \in \Real^2$,
with no-flux boundaries at
$\z=\zmin$ and $\z=\zmax$. Let
$\H(\x,\y)\equiv\zmax(\x,\y)-\zmin(\x,\y)$ and
$0<\H(\x,\y)\le\veps\ll1$. 
Then 
\(
\v(\x,\y,\z,\t) = \u(\x,\y,\t)+\O{\veps^2}       
\) ~\cite{epaps},
and Eq.~\eq{3D-pde} in the leading order in $\veps$ reduces to
the following 2D approximation:
\begin{align} \eqlabel{2.5D-pde}
\u_\t = \f(\u) + \D \frac{1}{\H(\x,\y)} \nabla\cdot\left( \H(\x,\y) \nabla\u\right)
  + \O{\veps^2} .
\end{align}

We rewrite Eq.~\eq{2.5D-pde} in the form
\begin{align} \eqlabel{2.5D-perturbed}
  \u_\t = \f(\u) + \D\nabla^2\u + \eps\h(\u,\x,\y,\nabla)
\end{align}
where
\begin{align}  \eqlabel{2.5D-perturbation}
  \eps\h = \eps\D\left(\nabla\K\right)\cdot\left(\nabla\u\right), \qquad \eps\K = \ln\H. 
\end{align}
Eqs. \eqtwo{2.5D-perturbed,2.5D-perturbation} will be treated as a perturbation
problem with the formal small parameter $\eps$,  distinct from
the small parmeter $\veps$. So, for Eq.~\eq{2.5D-perturbed} we assume existence of a rigidly rotating
spiral wave solution $\U$ at $\eps=0$. 

In what follows, we explicitly calculate scroll wave drift for
  three given geometries corresponding to abrupt changes in domain
  thickness, i.e. a thickness step, ditch, and circular bulge. First,
  we consider a step in thickness, as in \fig{step3d}: 
\begin{align}
  \H(\x,\y) =
  \begin{cases}
    \H_+, & \x>\xs, \\
    \H_-, & \x<\xs. \\
  \end{cases}
\end{align}
Since $\K=\K(\x)$, one has $\eps\h=\eps\D\K_\x\u_\x$.  
With $\Heav$ the Heaviside step function, we have 
$\eps\K=\ln\left(\H_-\right) + \eps\Heav(\x-\xs)$, 
$\eps=\ln(\H_+/\H_-)$, such that
\begin{align}
  \eps\h = \eps\dirac(\x-\xs)\D\u_\x .            \eqlabel{epsh}
\end{align}

Eqs.~(12,13,14) of \cite{Biktasheva-etal-2010-PRE} predict the drift
velocity $\eps F(\vR) = \eps\left(\Fx+\i\Fy\right)$ as overlap integrals of translational response functions:
\begin{align}
& \Df{\vR}{\t} = \eps\vF(\vR)=\eps\left(\Fx,\Fy \right),             \eqlabel{F1-rhs} \\
& \F(\vR) = 
  \int\limits_{0}^{\infty}\oint 
  \RFS\left(\crr,\cra\right)^\dagger\balp (\crr, \cra; \vR)
  \,\d\cra\,\crr\d\crr
  ,
                                                  \eqlabel{F1-inner} \\
& \balp (\crr, \cra; \vR) = 
        \oint
        \e^{-\i\phase} \; \tilde \h(\U,\crr,\cra,\phase) \, \frac{\d\phase}{2\pi},
                                                  \eqlabel{F1-balp}
\end{align}
where $\tilde\h$ is the perturbation $\h$, calculated for $\u=\U$ and
considered in the frame corotating with the spiral, $\RFS$ are
(translational) response functions of the
spirals, and $\dagger$ stands for conjugate transposed.
Counter-clockwise rotating
spirals and their
response functions calculated for the two selected models using
\DXSPIRAL~\cite{dxspiral,epaps} are illustrated in Fig.~5 in ~\cite{epaps};
change of chirality of the spirals corresponds to complex
  conjugation of $\RFS$, $\balp$ and $\F$.
Evaluation of the integral~\eq{F1-balp} with account of~\eq{epsh} and the
coordinate transformations $\h(\vR,\t) = \tilde\h(\crr,\cra,\phase)$,
$\vR=(\X,\Y)$, $\vr=(\x,\y)$, $\dists=\X-\xs$,
$\cra=\cma(\vr-\vR)+\phase$, $\crr=\cmr(\vr-\vR)$,
$\x+\i\y\equiv\cmr(\vr)\,\exp(\i\cma(\vr))$ gives
\begin{align}
    \balp 
    =
    \begin{cases}
      0, & \crr\le|\dists|, \\
      \dfrac{\D \e^{-\i\cra}}{\pi\sqrt{\crr^2-\dists^2}} 
      \left[
        \dfrac{\dists^2}{\crr^2}\,\U_\crr
        -\dfrac{\i(\crr^2-\dists^2)}{\crr^3}\,\U_\cra
      \right],
     & \crr>|\dists|.
    \end{cases}                                   \eqlabel{balp1-of-d}
\end{align}
Eqs.~\eq{F1-inner} and \eq{balp1-of-d} define 
the specific force produced by the thickness step which
depends only on the distance between the current spiral centre and the
step  and is an even function about the position of the step, 
\begin{align}
  & \F(\vR)=\Fs(\dists), \qquad \dists \equiv \X-\xs, \\
  & \Fs(-\dists)=\Fs(\dists)=\Fsx(\dists)+\i\Fsy(\dists). \eqlabel{Fs}
\end{align}

\sidefigure{fig2.eps}{%
  Asymptotic theory vs numerical simulations for interaction of a
  scroll wave with a thickness step, for the FitzHugh-Nagumo system
  (panels a-c) and the Oregonator model (panels d-f). %
  (a,d): Spiral wave snapshot (red color component: $u$
  field, green color component: $v$ field, blue
  color component: $H$ field), with the previous tip path
  (solid white line) starting at A, another path starting at a different
    point (B), and loci $\dists=\pm\distsc$
    (dashed yellow lines), in~2D system~\eq{2.5D-pde}. %
  (b,e): Components of the specific force $\Fs$~\eq{Fs} %
  calculated for counter-clockwise spirals. %
  (c,f): Drift  speed along the step in 3D
  system~\eq{3D-pde}, 2D system~\eq{2.5D-pde} and asymptotic predicted
  by~\eq{step-drift}. \\[0.5ex] %
}{asymp}

The components of the function $\Fs(\dists)$ for the two selected
models are shown in~\fig{asymp}(b,e). An important feature are zeros
of $\Fsx$ for $\dists=\pm\distsc$ in both models. Assuming without
loss of generality that $\xs=0$, the drift of a spiral
wave is then described 
asymptotically by
\begin{align}
  \Df{\X}{\t} = \eps\Fsx(\X),
  \quad
  \Df{\Y}{\t} = \eps\Fsy(\X),
  \quad
  \eps=\ln\left(\frac{\H_+}{\H_-}\right).         \eqlabel{step-ODE}
\end{align}
\Fig{asymp} illustrates predictions of the theory for the case of 
a thickness step and their
comparison with the direct
numerical simulations of both the 2D thickness-reduced
system~\eq{2.5D-pde} and the full 3D system~\eq{3D-pde}. Numerical
simulations for both selected models were done with
\BEATBOX~\cite{beatbox,epaps}.
The relevant attractor for~\eq{step-ODE} is
\begin{align}
\X=-\distsc, \qquad \Y=\Y_0+\eps\Fsy(-\distsc)\t,    \eqlabel{step-drift}
\end{align}
where $\Fsx(-\distsc)=0$, $\Fsx'(-\distsc)<0$. That is, in both models
the spirals attach to the step at its thinner side and drift along
with the speed $|\eps\Fsy(-\distsc)|$. The speed of the drift is
proportional to $\eps=\ln(\H_+/\H_-)$, and the direction of the drift
depends on the spiral chirality: compare~\fig{asymp} (a) and
(d).

As a second geometry, let us consider the following thickness profile: for some
$\xleft<\xright$, 
\begin{align}
  \H(\x,\y) =
  \begin{cases}
    \Hout, & \x<\xleft, \\
    \Hin, & \xleft<\x<\xright, \\
    \Hout, & \xright<\x, \\
  \end{cases}
\end{align}
which means a ``ridge'' for $\Hin>\Hout$ and a ``ditch'' for
$\Hin<\Hout$. This case is easily reduced to the previous because
\(
  \H(\x,\y) = \Hin +
  (\Hout-\Hin)\left(\Heav(\x-\xleft)-\Heav(\x-\xright)\right) ,
\)
hence the formal perturbation is
\(
  \eps\h =
  \eps\, \left[\dirac(\x-\xleft)-\dirac(\x-\xright)\right]\,\D\u_\x ,
\)
where $\eps=\ln(\Hin/\Hout)$. 
Let $\xleft=\xs-\width/2$, $\xright=\xs+\width/2$. 
We use the linearity of~\eq{F1-rhs}--\eq{F1-balp} and the
previous result to get the interaction force in the form
\begin{align}
  & 
  \F(\vR)=\Ft(\dists;\width)=-\Ft(-\dists;\width), \qquad \dists \equiv \X-\xs, 
  \\
  & 
  \Ft=\Ftx+\i\Fty=
  \Fs\left(\dists+\frac{\width}{2}\right)-\Fs\left(\dists-\frac{\width}{2}\right). \eqlabel{Ft}
\end{align}

\dblfigure{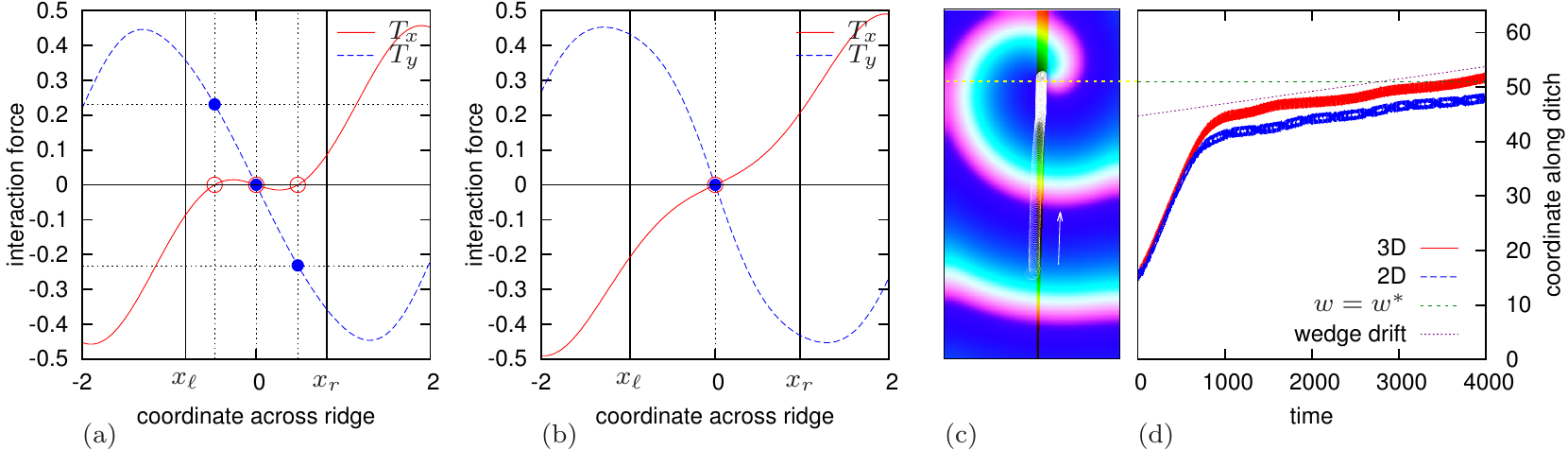}{%
  Results of asymptotic theory of interaction with a ridge/ditch, and
  comparison with simulations, for FitzHugh-Nagumo system~\eq{FHN}. %
  (a,b) Specific interaction force $\Ft(\dists)$ \eq{Ft} calculated for counter-clockwise spirals: %
  (a) for $\width=1.621<\widthc$; 
  (b) for $\width=1.953>\widthc$. 
  (c) Spiral wave snapshot (red color component: $u$ field, green
  color component: $v$ field, blue color component: $\H$ field), with
  the previous tip path (white line), drifting along a cuneiform
  ditch, of width $0.24$ 
  at the lower end, linearly growing to $2.16$ 
  at the upper end, box size $32\times64$, 
  and $\Hout/\Hin=1.2$. %
  (d) Coordinate of the spiral tip along the ditch as a function of
  time. %
  The horizontal dashed line shows location of
    the ditch width $\width=\widthc$ corresponding to the point of the
    pitchfork bifurcation of $\Ft(\dists;\width)$. %
  The slope of the dotted line represents the slow drift speed due to
  the sides of the ditch being non-parallel. %
}{trough}

\Fig{trough}(a,b) shows the components of
$\Ft(\dists;\width)$ for two selected values of the ridge width
$\width$, illustrating a pitchfork bifurcation of $\Ftx$ roots.
The bifurcation condition
$\Ftx(\dists;\width)=\@_\dists\Ftx(\dists;\width)=0$,
observation that the bifurcation happens at $\distsc=0$ 
and evenness of $\Fs(\dists)$ gives the critical value of the width
implicitly as the condition
$\Fsx'(\widthc/2)=0$. For the FHN system, there are two positive roots for $\Fsx'(\cdot)$
(see~\fig{asymp}(b)), the smaller giving
$\widthc\approx1.769$. 
For a ditch ($\eps<0$), this predicts neutrally
  stable equilibria along the middle line of the ditch if
  $\width>\widthc$ (\fig{trough}(b)), and a drift along either side of
  the ditch, in one direction or the other, depending on the initial
  condition, if $\width<\widthc$ (\fig{trough}(a)).

\Fig{trough}(c,d) illustrates the drift along a cuneiform ditch,
i.e. a ditch with almost constant but slowly varying
width. The coordinate scale along the ditch is
  the same in both panels, with the bifurcation width $\widthc$
   designated by the dashed horizontal line. %
We see that below this line the spiral wave drifts in accordance with
the theory  and slows down markedly in the
vicinity of this line. It does not stop completely but proceeds
further, albeit at a much slower speed, which may be seen as a
  transient pinning. This slow drift is due to the ``wedging'' effect of the
varying width: at $\width\ge\widthc$, the forces from the two opposite
steps, constituting the banks of the ditch, do not compensate each other
exactly due to the angle between them. To estimate roughly the
associated correction, let the wedge angle be $\wedge\ll1$. Then the
wedge-forced component of the drift speed at the bifurcation point is
$2\eps\Fsx(\widthc)\sin(\wedge/2)\approx\eps\Fsx(\widthc)\wedge$.  For
the simulation shown in \fig{trough}(c,d), we have
$\wedge\approx0.03$, and $\Fsx(\widthc)\approx0.4142$, hence the drift
speed
$\eps\wedge\Fsx(\widthc)\approx0.002266$.  This wedge-forced drift speed is
represented by the dotted line in~\fig{trough}(d) and
agrees well
with the simulations. %
  If the initial position of the spiral is where $\width\gtrsim\widthc$,
  then it undergoes only the slow, wedge-forced drift from the start
  (not shown).

\sglfigure{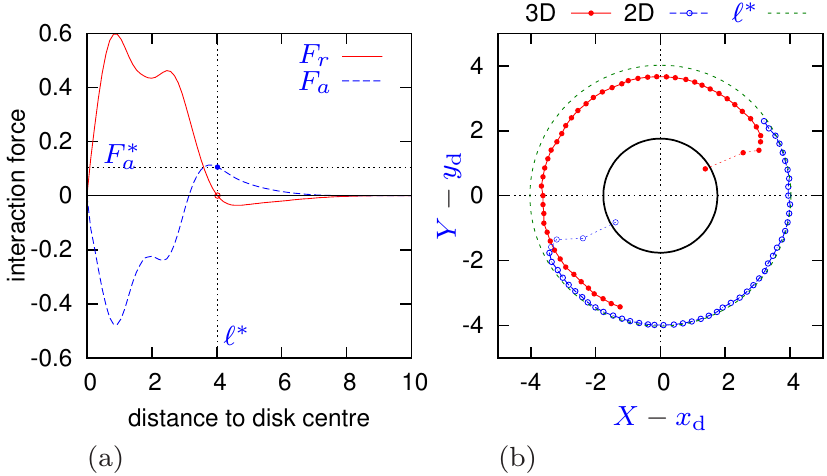}{%
  Interaction of a spiral with a disk-shape bulge in Oregonator
  model. %
  (a) Components of the interaction force calculated according to
  \eq{F1-inner},\eq{F1-balp},\eq{balp1-of-ell},\eq{Fdr-Fda}, for
  $\Ri=225/1280\approx1.756$. %
  (b) Tip trajectories in simulations of duration corresponding to
  half of predicted orbiting period (lines as indicated by the
  legend), together with initial transients (thin dotted lines). The
  green dashed circle: the theoretically predicted stationary orbit
  of the spiral
  centre drift. The black solid circle: the boundary of the bulge. %
}{disk}

Drift caused by ridge or ditch features may help to
  understand dynamics of scroll waves in atrial geometry, say around
  pectinate muscles~\cite{Yamazaki-etal-2012-CVR}. Another feature,
  specifically analyzed in~\cite{Yamazaki-etal-2012-CVR} by numerical
  simulations was a circular bulge. To see what our theory can say
  about that, 
let us consider thirdly a thickness perturbation of the form
\begin{align*}
  \H(\x,\y) = \H_0\left( 
    1 + \eps\,\Heav\left(\Ri^2-(\x-\xd)^2-(\y-\yd)^2\right)
  \right)
\end{align*}
i.e. thickening (for $\eps>0$) or thinning (for $\eps<0$) in a
disk-shaped area of radius~$\Ri$. Then we have
\begin{align}
  \balp =
  \dfrac{\e^{\i\angled}\e^{-\i\cra}\D}{\pi \crr \distd
    \sqrt{1-\cocl^2}} \left[ (\distd\cocl^2+\crr\cocl)\,\U_{\crr}
    - \dfrac{\i\distd (1-\cocl^2)}{\crr}\,\U_{\cra} \right]
  \eqlabel{balp1-of-ell}
\end{align}
for $\r\in(|\Ri-\distd|,\Ri+\distd)$, and $\balp=0$ otherwise. 
Here $\distd\e^{\i\angled}=(\xd-\X)+\i(\yd-\Y)$ represents the vector
from the current spiral centre $(\X,\Y)$ to the bulge
centre $(\xd,\yd)$, and 
$\cocl =
\left(\Ri^2-\distd^2-\crr^2\right)/\left(2\crr\distd\right)$. 
Hence the interaction force is
\begin{align}
  \Fx+\i\Fy = \e^{\i\angled}\left(\Fdr(\distd)+\i\Fda(\distd)\right). 
                                                  \eqlabel{Fdr-Fda}
\end{align}
The radial $\Fdr(\distd)$ and the azimuthal $\Fda(\distd)$ components
calculated for the Oregonator model~\eq{BZO} for an arbitrarily chosen
disk radius $\Ri$ are shown in \Fig{disk}(a). We observe there is a root of
$\Fdr(\distd)$ at $\distd=\distdc\approx4.023$ and the corresponding
value of the specific force $\Fdac=\Fda(\distdc)\approx0.1055$  predicts long-term
behaviour of a spiral starting from an appropriate initial condition
as ``meander'' or ``orbital movement'' along a circle of radius
$\distdc$ with the linear speed $\eps\Fdac$, and an orbit period
of $2\pi\distdc/(\eps\Fdac)\approx1314$. \Fig{disk}(b) compares these
predictions with results of 2D and 3D numerical simulations at
$\eps=\log(1.2)$. This result is similar to the case considered
phenomenologically 
in~\cite{Yamazaki-etal-2012-CVR} and is analogous to ``orbital
motion'' described in~\cite{Biktashev-etal-2010-PRL} for localized
parametric heterogeneities.


To summarise, the movement of transmural scroll waves through thin layers
of excitable media of varying thickness can be approximately described
by thickness-averaged two-dimensional equations, and a corresponding
2D perturbation theory can be successfully applied within its
limits. Our theory shows the propensity of scrolls to interact with
sharp features of the layer geometry. In the examples
  considered, this interaction caused a scroll to position itself at a
  certain distance from a sharp feature and drift along/around it with
  the speed determined by the feature's magnitude, measured by the
  relative variation of the thickness. This is distinct from and not
reducible to previously known geometric effects such as filament
tension or curvature-induced drift, and is completely independent from
other factors that may cause drift, such as
  parametric inhomogeneities or external forcing (see
  e.g.~\cite{Biktasheva-etal-2010-PRE}).  Interaction with sharp
features can manifest nontrivial attractor structures, depending on
the geometric parameters.  These predictions should be immediately
testable in experiments with the Belousov-Zhabotinsky reaction,
can be used in experiments, say for precision
  positioning of scrolls, 
and may have important implications for understanding the evolution of
re-entrant waves of excitation in the heart, particularly in
atria which have an abundance of geometric features. For
  instance, our results give a theoretical explanation, and hence
  suggest a universal character  of scroll wave ``anchoring'' and
  ``meandering'' caused by thickness variations, which are implicated
  in perpetuation of atrial
  fibrillation~\cite{Yamazaki-etal-2012-CVR}. 


\textbf{Acknowledgments} H.D. is supported by FWO-Flanders (Belgium).
Development of \DXSPIRAL\ and \BEATBOX\ was supported by EPSRC grants
EP/D074789/1 and EP/I029664/1 (UK).

\clearpage
\+{\Ar}{A}                      
\+{\a}{a}                       
\+{\b}{b}                       
\+{\C}{C}                       
\+{\G}{G}                       
\+{\iset}{I}                    
\+{\j}{j}                       
\+{\k}{k}                       
\+{\LH}{L}                      
\+{\lam}{\lambda}               
\+{\n}{n}                       
\+{\nvec}{\vec{n}}              
\+{\Rad}{R}                     
\+{\rh}{\rho}                   
\+{\rhx}{\xi}                   
\+{\rhy}{\eta}                  
\+{\rhz}{\zeta}                 
\+{\rmax}{r_{\max}}             
\+{\rmid}{r_{\textrm{mid}}}     
\+{\rmin}{r_{\min}}             
\+{\s}{\sigma}                  
\+{\T}{T}                       
\+{\ut}{\hat\u}                 
\+{\vv}{\mx{g}}                 

\+{\ch}{\chi}                   
\+{\dr}{\Delta\rho}             
\+{\dt}{\Delta t}               
\+{\dx}{\Delta x}               

\begin{widetext}
\begin{centering}
{\large\bf Supplementary material:} \\
{\large\bf Drift of scroll waves in thin layers caused by thickness features} \\
I.V.~Biktasheva, H.~Dierckx, V.N.~Biktashev \\
\end{centering}

\subsection{Thin layer asymptotics}


When the thickness of the excitable medium layer is much smaller than
the diffusion length $\sqrt{||\D||/\max\left(||\@\f/\@\v||\right)}$,
then we can expect the concentration field $\v$ to be nearly constant
across the thickness of the layer, thus being effectively a
two-dimensional field.  %
  This is, of course, ``intuitively obvious'', and simple considerations
  based on conservation of matter can help to immediately
  ``guess'' the resulting equation~\eq{2.5D-pde}. We, however, would
  like to also have some estimate of the accuracy of that reduced
  equation, in order to know its limitations, so we feel that a
  simple guess is not quite sufficient. Problems for partial
  differential equations posed in thin or slender domains 
  occur in many fields of applied mathematics and there are many
  works considering such problems in various situations and
  with various practical purposes. In this respect we mention two
  examples: 
  E. Yanagida, ``Existence of stable stationary solutions of scalar
  reaction-diffusion equations in thin tubular domains'',
  \textit{Applicable Analysis} \textbf{36}:171-188 (1990), %
  and %
  J. K. Hale and G. Raugel, ``Reaction-diffusion equations on thin
  domains'', \textit{J. Math. pures et appl.} \textbf{71}:33--95
  (1992), %
  in which the reduction has been done rigorously and in the context
  of reaction-diffusion equations, making them particularly close to
  what we require for our study. However those rigorous results have been
  obtained under certain assumptions which make them technically
  inapplicable to our case. So, Yanagida (1990) writes about
  ``tubular'' domains, that is, there is only one ``long''
  dimension. It is also focused on studying stationary solutions, and
  presented for the case of one-component reaction-diffusion
  system. All of these assumptions are essential for their results,
  and all of them are unsuitable for us, as the spiral wave solutions
  require two spatial dimensions, at least two components and are
  non-stationary. The work by Hale and Raugel (1992) goes further in
  that it allows up to two long dimensions and considers
  non-stationary solutions, but it is still restricted to
  one-component reaction-diffusion systems with extra requirements on
  the kinetic term. We are not aware of either rigorous or formal
  results that would quite cover our needs, hence we present our own
  derivation, even though it is only a formal asymptotic. %

To exclude the effects of the
curvature we assume the layer to be flat on the macroscopic scale.  The formal
setup is as follows:
\begin{align*}
  & \@_\t\v = \D\nabla^2\v + \f(\v), 
  \qquad (\x,\y)\in\Real^2,
  \qquad \z\in(\zmin(\x,\y;\veps),\zmax(\x,\y;\veps),
  \\
  & 
  \zmin(\x,\y;\veps) = 0, 
  \qquad
  \zmax(\x,\y;\veps)=\H(\x,\y;\veps)=\veps\Ht(\x,\y) , 
  \qquad
  \qquad \veps\ll 1,
  \\
  & \nvec(\zmin)\cdot\D\nabla\v(\x,\y,\zmin) = 0, \\
  & \nvec(\zmax)\cdot\D\nabla\v(\x,\y,\zmax) = 0,
\end{align*}
where $\nvec(\cdot)$ is the normal vector at the corresponding
surface. 
The shape of the layer is asymmetric and it may seem that variations of
thickness may introduce small curvature effects; however it is easy to
see that the above formulation is exactly equivalent to a symmetric
one, 
\begin{align*}
  \zmin(\x,\y;\veps) = -\veps\Ht(\x,\y), \qquad \zmax(\x,\y;\veps)=\veps\Ht(\x,\y).
\end{align*}

The boundary conditions at $\z=\zmin$, $\z=\zmax$ mean that the
flux lines for the $\v$ concentrations need to intersect the domain
boundary perpendicularly. To accommodate this property in our
asymptotic solution, we switch from the original Cartesian coordinates
$\r=(\x^\j)=(\x,\y,\z)$ to new curvilinear coordinates
$(\rh^\j)=(\rhx,\rhy,\rhz)$, $\j= 1,2,3$, in the following way:
\begin{itemize}
\item Coordinate $\rhz$ is ``transmural'', that is
  \begin{align}
    \z(\rhx,\rhy,0)=\zmin(\x(\rhx,\rhy,0),\y(\rhx,\rhy,0)),
    \qquad
    \z(\rhx,\rhy,1)=\zmax(\x(\rhx,\rhy,1),\y(\rhx,\rhy,1)).
    \eqlabel{curvicoord-transmural}
  \end{align}
\item The other two ``intramural'' coordinates $(\rhx,\rhy)$ are
  chosen locally orthogonal to $\rhz$, i.e.
  \begin{align}
    \df{\r}{\rhx}\cdot\df{\r}{\rhz}=\df{\r}{\rhy}\cdot\df{\r}{\rhz}=0.
    \eqlabel{curvicoord-orthogonal}
  \end{align}
\item The intramural coordinates match the horizontal Cartesian
  coordinates in the sense that 
  \begin{align}
    \x(\rhx,\rhy,0)=\rhx, \qquad
    \y(\rhx,\rhy,0)=\rhy .
    \eqlabel{curvicoord-intramural}
  \end{align}
\end{itemize}
Thus the choice of the curvilinear coordinates is fully determined by
the choice of function $\rhz(\r)$. A convenient choice is ``heat
coordinates'', when 
$\rhz(\r)=\T(\r;\veps)$ which is a solution of the boundary-value problem
\begin{align}
  & \nabla^2\T(\x,\y,\z)=0, \qquad \z\in(0,\veps\Ht(\x,\y));
  \nonumber\\
  & \T(\x,\y,0) = 0, 
  \nonumber\\
  & \T(\x,\y,\veps\Ht(\x,\y)) = 1,                 \eqlabel{Tproblem}
\end{align}
i.e. it is identified with an established temperature distribution when a
unit temperature drop is imposed across the layer. Then the lines
$\rhx=\const$, $\rhy=\const$ can be interpreted as the lines of heat
flux, and in the new coordinates, the boundary conditions become
simply condition of zero derivative in $\rhz$, as the vectors $\nvec$
are tangent to these flux lines. 
The two leading terms of the asymptotic of the solution
of~\eq{Tproblem} are 
\begin{align*}
  \T = \frac{\z}{\veps\Ht} 
  + \frac{\veps^2\Ht^2}{6}
  \left(1-\frac{\z^2}{\veps^2\Ht^2}\right)\frac{\z}{\veps\Ht} \, 
  \left(\left(\nabla\LH\right)^2-\nabla^2\LH\right)
  + \O{\veps^4};
  \qquad 
  \left( \z\in[0,\veps\Ht] \right)
\end{align*}
where
\begin{align*}
  \Ht=\Ht\left(\x,\y\right), \qquad  \LH=\LH(\x,\y)=\ln\Ht(\x,\y) .
\end{align*}

Let $\g_{\j\k}$ be the metric tensor in the curvilinear coordinates
$\rh^\j$, that is 
\begin{align*}
  \g_{\j\k} = \df{\r}{\rh^\j} \cdot \df{\r}{\rh^\k} = \g_{\k\j}. 
\end{align*}
Due to the local orthogonality condition~\eq{curvicoord-orthogonal}, 
we have automatically $\g_{13}=\g_{23}=0$, and the metric tensor takes
the block-diagonal form
\begin{align*}
  \Mx{ \g_{\j\k} } = \Mx{
    \gg_{11} & \gg_{12} & 0 \\
    \gg_{21} & \gg_{22} & 0 \\
    0 & 0 & \g_{33}
  }
\end{align*}
To find the asymptotics of the remaining components of the metric
tensor, we first find $\r(\rhz)$ as a solution of a Cauchy problem
for the ODE system, defined by function $\T(\r)$ found previously and
depending on $\rh^{1,2}$ as parameters,
\begin{align}
  & \df{\r}{\rhz}=\lam(\rhz)\nabla\T(\r) \\
  & \T(\r(\rh^\j))\equiv\rhz, \\
  & \r(0)=(\rhx,\rhy,0). 
\end{align}
The asymptotics for the solution are
\begin{align*}
  & \lam = \veps^2 \Ht^2(\rhx,\rhy) + \O{\veps^4}, \\
  & \x = \rhx-\frac12\veps^2\Ht\Ht_\x\rhz^2+\O{\veps^4}, \\
  & \y = \rhy-\frac12\veps^2\Ht\Ht_\y\rhz^2+\O{\veps^4}, \\
  & \z = \veps\Ht\,\rhz
  - \frac16\veps^3 \Ht^3 \left[
    \left(\nabla\LH\right)^2 \left(\rhz + 2\rhz^3\right)
    -\nabla^2\LH\left(\rhz-\rhz^3\right)
  \right]
  + \O{\veps^5},
\end{align*}
where $\Ht$, $\LH$ and their derivatives are evaluated at
$(\rhx,\rhy)$. This then leads to the asymptotics of the
components of the metric tensor in the form
\begin{align*}
  & 
  \gg_{\a\b}=\delta_{\a\b}+\O{\veps^2},
  \qquad \a,\b=1,2, 
  \\ 
  &  
  \g_{33} = \veps^2 \Ht^2 - \frac13 \veps^4\Ht^4 \left[
    \left(\nabla\LH\right)^2\left(1+3\rhz^2\right)
    -\nabla^2\LH \left(1-3\rhz^2\right)
  \right]
  +\O{\veps^6}.
\end{align*}
Here for $\g_{33}$ we go beyond the leading term; the reason for that
will become clearer later.

For the curvilinear Laplacian we shall also need
\begin{align*}
  |\g| =\det\Mx{\g_{\j\k}}=\g_{33}\det\Mx{\gg_{\a\b}} = \veps^2 \Ht^2 + \O{\veps^4}, 
\end{align*}
and
\begin{align*}
  \Mx{ \g^{\j\k} } 
  = \Mx{ \g_{\j\k} }^{-1}
  = \Mx{ 
    \Mx{\gg_{\a\b}} & \mx{0} \\
    \mx{0}\Tr & \g_{33}
  }^{-1}
  = \Mx{ 
    \Mx{\gg_{\a\b}}^{-1} & \mx{0} \\
    \mx{0}\Tr & \g_{33}^{-1}
  }
  = \Mx{ 
    \Mx{\gg^{\a\b}} & \mx{0} \\
    \mx{0}\Tr & \g^{33}
  } ,
\end{align*}
where
\begin{align*}
  \gg^{\a\b} = \delta^{\a\b}+\O{\veps^2},
\end{align*}
and
\begin{align*}
  \g^{33} = \frac{1}{\veps^2\Ht^2} + \frac13 \left[
      \left(\nabla\LH\right)^2\left(1+3\rhz^2\right)
      -\nabla^2\LH \left(1-3\rhz^2\right)
    \right]
    +\O{\veps^2} .
\end{align*}
Here we see the reason for a higher accuracy: the second term of the
asymptotic for $\g^{33}$ has order $\O{1}$ so omitting it could have
affected our main result.

Now we are ready to calculate the Laplacian of the concentration
field, which in the new curvilinear coordinates is represented by the
Laplace-Beltrami operator (using Einstein summation convention)
\begin{align*}
  \nabla^2\v = |\g|^{-1/2}\df{}{\rh^\j}\left( |\g|^{1/2} \g^{\j\k}
    \df{\v}{\rh^k} \right).
\end{align*}

Taking into account the results obtained above, this works out to be
\begin{align*}
  \nabla^2\v &
  = \frac{1}{\veps^2\Ht^2} \v'' + \nabla\LH\nabla\v + \nabla^2\v + \G \, \v'' + \O{\veps^2}
\end{align*}
where
\begin{align*}
  \G=\G(\rhx,\rhy,\rhz)
  =\left(\frac13+\rhz^2\right)\left(\nabla\LH\right)^2
  -\left(\frac13-\rhz^2\right)\nabla^2\LH .
\end{align*}
On the right-hand side, functions $\Ht$, $\LH$ and their derivatives
are evaluated at $(\rhx,\rhy)$, 
$\nabla$ is the gradient operator in the $(\rhx,\rhy)$ plane and the
prime $'$ stands for differentiation by $\rhz$. 

The reaction-diffusion equation in the new coordinates then takes the form
\begin{align*}
  \@_\t\v = \D\left[
    \frac{1}{\veps^2\Ht^2} \v'' + \nabla\LH\nabla\v + \nabla^2\v + \G \, \v'' + \O{\veps^2}
  \right]
  + \f(\v)
\end{align*}
%
with the boundary conditions 
\begin{align*}
  \D\@_3\v(\rhx,\rhy,0,\t)=\D\@_3\v(\rhx,\rhy,1,\t)=0.
\end{align*}
We shall look for the solution of this problem in the form of an
asymptotic series in $\veps^2$, and consider the two leading terms,
\begin{align*}
  \v(\rhx,\rhy,\rhz,\t;\veps) = \u(\rhx,\rhy,\rhz,\t) +
  \veps^2\vv(\rhx,\rhy,\rhz,\t)
  +\O{\veps^4}.
\end{align*}
%
Upon rewriting our problem in the form
\begin{align*} &
  \D\left(\u+\veps^2\vv\right)'' = \veps^2\Ht^2 \left[
    \@_\t\left(\u+\veps^2\vv\right) 
    - \frac{1}{\Ht}\D\nabla\left(\Ht\nabla\left(\u+\veps^2\vv\right)\right)
    -\G\D\left(\u''+\veps^2\vv''\right)
    - \f\left(\u+\veps^2\vv\right)
  \right] + \O{\veps^4},
  \\ & 
  \left[\D(\u+\veps^2\vv)'\right]_{\rhz=0,1}=\O{\veps^4} ,
\end{align*}
we get in the order $\O{1}$ 
\begin{align*}
  \D\u''=0 ,
  \qquad
  \left[\D\u'\right]_{\rhz=0,1}=0 ,
\end{align*}
wherefrom $\u'=\@_\rhz\u\equiv0$, i.e. function $\u$ depends only on
$\rhx,\rhy$ and $\t$ but not on $\rhz$, as expected. 
Further, in
the order $\O{\veps^2}$ we have
\begin{align*}
  \D\vv'' = \Ht^2 \left[
    \@_\t\u 
    - \frac{1}{\Ht}\D\nabla\left(\Ht\nabla\u\right)
    - \D\G\u''
    - \f(\u)
  \right] ,
  \qquad
  \left[\D\vv'\right]_{\rhz=0,1}=0 .
\end{align*}
This is a two-point ODE boundary-value problem for $\vv(\rhz)$
depending on $\rhx,\rhy$ and $\t$ as parameters.  Note that the term
$\D\G\u''$ vanishes by the $\O{1}$ result, and the remaining free term
is a constant, i.e. does not depend on $\rhz$.
This problem is solvable if this constant vanishes, i.e. 
\begin{align*}
  \@_\t\u - \frac{1}{\Ht}\D\nabla\left(\Ht\nabla\u\right) - \f(\u) 
  = 0
\end{align*}
which gives the leading term for the equation~\eq{2.5D-pde} of the
main text, since $\nabla\Ht/\Ht=\nabla\H/\H$.

The higher-order approximations cannot be obtained within the same
asymptotic procedure, and instead one would need to look for the
asymptotic expansion of the right-hand side of the evolution equation
for $\u$, that is, $\@_\t\u=\sum_\n\veps^{2\n}\ut_\n(\u,\H,\nabla)$,
i.e. admit different asymptotic orders in the right-hand side of the
``master equation''. Such asymptotic technique is outlined e.g. in
V.N. Biktashev, ``Envelope Equations for Modulated Non-conservative
Waves'', in \textit{IUTAM Symposium Asymptotics, Singularities and
  Homogenisation in Problems of Mechanics}, pp. 525-535, ed. by
A.B. Movchan, Kluwer, Dordrecht-Boston-London, 2003,
\url{http://empslocal.ex.ac.uk/people/staff/vnb262/publ/iutam-2002/index.html}.
We do not present the resulting derivation here as it
  is tedious and its precise results are not required in the main part
  of the paper; however the most important fact is that the
  next-to-leading order terms in the evolution equation are
  $\O{\veps^2}$, and this fact is already evident from the above. 

The above derivation was done under the assumption of smoothness of the
thickness profile $\Ht(\x,\y)$. Hence the applications considered in
the main paper will be formally covered by this approximation and the
2D spiral perturbation theory, if the ``sharp features'' considered
there are smooth on the scale of $\Ht$ but sharper than the typical
scale of the effective response functions' support. Deviation from
this condition in actual 3D simulations may account for some of the
discrepancies between the 3D and 2D simulations.

\subsection{Response functions quadratures}

\subsubsection{Straight step}

The function~\eq{balp1-of-d} has a singularity at $\crr=|\dists|$, so
the resulting integral by $\crr$ cannot be adequately evaluated by the
usual trapezoidal rule.  So we proceed instead in the following
way. Let the radii grid be $\crr\in\{\j\dr\;|\;\j=0,1,2\dots\}$, and
$|\dists|=\k\dr$ for some $\k\in\Z_+$. Then, for a
regular function $\fun(\crr)$ and a constant $\s>-1$, we can write
%
\begin{align*}
  \int\limits_{|\dists|}^{\infty} \fun(\crr)(\crr^2-\dists^2)^\s \,\d\crr
  = \int\limits_{|\dists|}^{\infty} \Fun(\crr)(\crr-\dists)^\s \,\d\crr
  =
  \sum\limits_{\j=\k}^{\infty}
  \int\limits_{\j\dr}^{(\j+1)\dr} \Fun(\crr) (\crr-\dists)^\s \,\d\crr
  \approx 
  \sum\limits_{\j=\k}^{\infty} \C_\j \fun_\j \dr,
\end{align*}
where $\Fun(\crr)=\fun(\crr)(\crr+\dists)^\s$,
$\fun_\j=\fun(\j\dr)$, and linear interpolation of $\Fun(\crr)$ within
each subinterval gives
\begin{align*}
  \C_\j=
  \begin{cases}
    \frac{\dr^{2\s}}{2(\s+1)} (2\k)^\s, & \j=\k, \\
    \frac{\dr^{2\s}}{2(\s+1)}
    \left[(\j-\k+1)^{\s+1}-(\j-\k-1)^{\s+1} \right] (\j+\k)^\s,
    & \j>\k .
  \end{cases}
\end{align*}
For $\s=-1/2$, 
\begin{align}
  \C_\j=
  \begin{cases}
    0, & \j<\k , \\
    \frac{1}{\dr\sqrt{2\k}}, & \j=\k , \\
    \frac{1}{\dr\sqrt{\j+\k}}
    \left[\sqrt{\j-\k+1}-\sqrt{\j-\k-1}\right],
    & \j>\k .
  \end{cases}                                     \eqlabel{quadcorr}
\end{align}
In other words, we can use the usual trapezoidal formula, but should
multiply $\fun_\j=\fun(\j\dr)$ by coefficients $\C_\j$ given above
instead of 
$(\crr^2-\dists^2)^{-1/2}=(\crr^2-\dists^2)^\s=(\j^2-\k^2)^\s\dr^{2\s}=(\j^2-\k^2)^{-1/2}\dr^{-1}$. 

\subsubsection{Circular step}

Similarly, the quadrature for interaction with a disk involves $\balp$
described by~\eq{balp1-of-ell}, and so is also singular, as it
contains denominator $\sqrt{1-\cocl^2}$ which becomes zero at both
ends of the integration interval:
\begin{align*}
  \sqrt{1-\cocl^2} = \dfrac{1}{2\crr\distd}
  \left[ (\crr-\rmin)(\rmax-\crr)
    (\crr+\rmin)(\crr+\rmax)\right]^{1/2}
\end{align*}
where $\rmin=|\Ri-\distd|$, $\rmax=|\Ri+\distd|$. Doing as before, we get
\begin{align*}
  \int\limits_{\rmin}^{\rmax}
  \frac{\Fun(\crr)}{\sqrt{(\crr-\rmin)(\rmax-\crr)}} \,\d{\crr}
  =
  \sum\limits_{\j=0}^{\N} \C_\j \Fun(\crr_\j) \,\dr,
\end{align*}
where
\begin{align*}
  & \C_0 = \frac{1}{\dr^2} \left[
    (\Ar_{0}-\Ar_{1})\left(\crr_{1}-\rmid\right)
    + \Rad_{0}-\Rad_{1} \right], \\
  & \C_{\j}= \frac{1}{\dr^2} \left[
    \left(\Ar_{\j}-\Ar_{\j+1}\right)\left(\crr_{\j+1}-\rmid\right)
    +\left(\Ar_{\j}-\Ar_{\j-1}\right)\left(\crr_{\j-1}-\rmid\right)
    + 2\Rad_{\j} - \Rad_{\j+1} - \Rad_{\j-1} \right], 
  \; \j=1,\ldots,\N-1, \\
  & \C_{\N} =\frac{1}{\dr^2} \left[
    (\Ar_{\N}-\Ar_{\N-1})\left(\crr_{\N-1}-\rmid\right)
    - \Rad_{\N-1}+\Rad_{\N} \right], \\
  & \rmid=\frac12\left(\rmin+\rmax\right), \\
  & \dr=(\rmax-\rmin)/\N, \\
  & \r_\j=\rmin+\j\dr, \qquad \j=0,\ldots,\N, \\
  & \Rad_\j= \sqrt{(\rmax-\crr_\j)(\crr_\j-\rmin)}, \\
  & \Ar_\j=\arcsin\left(\dfrac{2(\crr_j-\rmid)}{\rmax-\rmin}\right).
\end{align*}

\subsection{Discretization}

\subsubsection{Two-dimensional simulations}

We use explicit Euler timestepping with time step $\dt$ and central
differencing for the diffusion term in~\eq{2.5D-pde}, with the following discretization
scheme
\begin{align*}
  \left[ \frac{1}{\Ht} \nabla\cdot\left(\Ht\nabla\u \right) \right]_{\ix,\jx}
  = \frac{1}{2\dx^2} \frac{1}{\Ht_{\ix,\jx}} 
  \sum\limits_{(\ii,\ji)\in \iset}
    \left(\Ht_{\ix+\ii,\jx+\ji}+\Ht_{\ix,\jx}\right) \,
    \left(\u_{\ix+\ii,\jx+\ji}-\u_{\ix,\jx}\right)
\end{align*}
where $(\ix,\jx)$ are 2D indices of the regular space grid of the size
$\Nx\times\Ny$ with step $\dx$ and $\iset=\{ 
(-1,0), 
(1,0), 
(0,-1),
(0,1) 
\}$. We employ standard non-flux boundary conditions.
The discretization parameters used for different results are described
in~\tab{discretization2D}.

\begin{table}[htbp]
\begin{tabular}{|l|l|l|l|l|}\hline
Figure                    & $\dt$       & $\dx$       & $\Nx$ & $\Ny$ \\\hline
\figref{asymp}(a,c)       & $6.4\E{-4}$ & $8\E{-2}$   & 400   & 400   \\\hline
\figref{asymp}(d,f)       & $0.25$      & $1.5\E{-3}$ & 200   & 200   \\\hline
\figref{trough}(c,d)      & $6.4\E{-4}$ & $8\E{-2}$   & 400   & 800   \\\hline 
\figref{disk}(b)          & $0.25$      & $1.5\E{-3}$ & 200   & 200   \\\hline
\end{tabular}
\caption[]{Discretization parameters in 2D simulations}
\tablabel{discretization2D}
\end{table}

\subsubsection{Three-dimensional simulations}

The discretization in 3D is a natural extension of the 2D scheme,
except instead of a fancy diffusion operator of~\eq{2.5D-pde} we now
have the plain diffusion of~\eq{3D-pde}. The complication now comes
from the more complicated geometry of the domain, which requires
special attention to the boundary conditions. We have employed the
following discretization:
\begin{align*}
  \left[ \nabla^2\v \right]_{\ix,\jx,\kx} 
  = \frac{1}{\dx^2} 
  \sum\limits_{(\ii,\ji,\ki)\in\iset}
  \ch_{\ix+\ii,\jx+\ji,\kx+\ki}
    \left(\v_{\ix+\ii,\jx+\ji,\kx+\ki}-\v_{\ix,\jx,\kx}\right)
\end{align*}
where $\ch_{\ix,\jx,\kx}=1$ if the grid point $\left(\ix,\jx,\kx\right)$
is within the domain and $0$ otherwise, and the neighbourhood template
is $\iset=\{ 
(-1,0,0), 
(1,0,0), 
(0,-1,0),
(0,1,0),
(0,0,-1),
(0,0,1)
\}$.
The space grid is regular with
step $\dx$, rectangular $\Nx\times\Ny$ in the horizontal direction,
and with $\kx\in\left\{1,\ldots,\Nz(\ix,\jx)\right\}$ where $\Nz(\ix,\jx)$
represents the thickness profile. In all our examples, $\Nz(\ix,\jx)$
takes only two values, denoted as $\Nzone$ and $\Nztwo$. 
The discretization parameters used for different results are described
in~\tab{discretization3D}.

\begin{table}[htbp]
\begin{tabular}{|l|c|c|c|c|c|}\hline
Figure              & $\dt$       & $\dx$     & $\Nx$ & $\Ny$ & 
$\Nzone/\Nztwo$ 
\\\hline
\figref{step3d}     & $6.4\E{-4}$ & $8\E{-2}$ & 400   & 400   &
$20/40$
\\\hline
\figref{asymp}(c)   & $6.4\E{-4}$ & $8\E{-2}$ & 400   & 400   &
$1/2,2/3,2/4,3/4,3/5,4/5,4/6,6/7,9/10,14/15,18/20,19/20,38/40,40/80,76/80$
\\\hline
\figref{asymp}(f)   & $1.5\E{-3}$ & $0.25$    & 200   & 200   &
$1/2,2/4,3/5,4/6,9/10,18/20,38/40,76/80,2/3,3/4,4/5,6/7,14/15,19/20,40/80$
\\\hline
\figref{trough}(d)  & $6.4\E{-4}$ & $8\E{-2}$ & 400   & 800   & 
$5/6$
\\\hline 
\figref{disk}(b)    & $1.5\E{-3}$ & $0.25$    & 200   & 200   & 
$5/6$
\\\hline
\end{tabular}
\caption[]{Discretization parameters in 3D simulations}
\tablabel{discretization3D}
\end{table}

\subsubsection{Response function computations}

\sidefigure{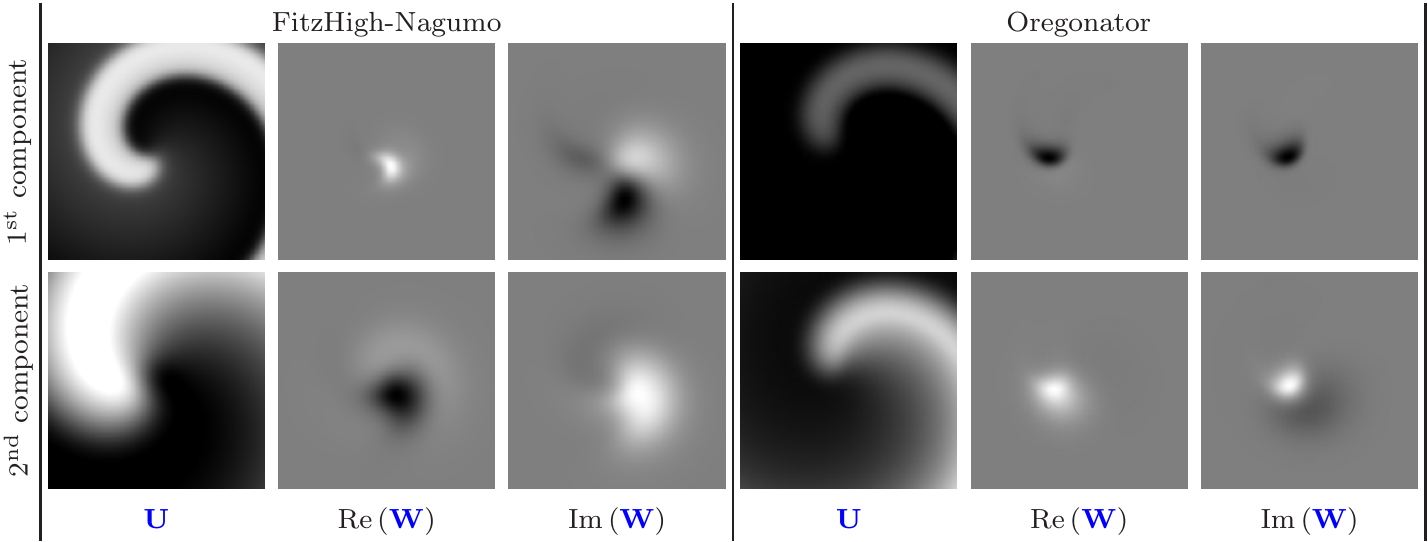}{%
  Density plots of the spiral waves $\U$ and response functions $\RFS$
  of the FitzHugh-Nagumo system~\eq{FHN} (central fragment
  $12.5\times12.5$) and Oregonator model~\eq{BZO} (central fragment
  $15\times15$). 
  Grey periphery of $\RFS$ components corresponds to zero, i.e. $\RFS$
  are well localized ensuring convergence of integrals~\eq{F1-inner}.
}{rfs}

For \DXSPIRAL\ computations of the FitzHugh-Nagumo model, we use disk
radius $\rp=25$, number of radial intervals $\Nr=1280$ and number of
azimuthal intervals $\Nt=64$. For the Oregonator model, we have
correspondingly $\rp=15$, $\Nr=128$ and $\Nt=64$. Density plots of the
spiral wave solutions and the corresponding response functions are
illustrated in~\fig{rfs}.

\subsection{Initial conditions}

We initiated a spiral wave in a large square domain using the phase
distribution method, as described e.g. in V.N. Biktashev, A.V. Holden
``Re-entrant waves and their elimination in a model of mammalian
ventricular tissue'' \textit{Chaos}, \textbf{8}(1):48--56 (1998), with
the centre of the Archimedian spiral phase distribution at the centre
of the square. A spiral wave initiated in that way was allowed to
fully establish itself during a few rotations without any perturbations, and
then saved to a disk file, to be used as an initial condition in 2D
simulations, cutting its 2D domain to size as appropriate. The scroll
waves for 3D simulations were obtained from the same spiral wave, by
extending it in the $z$ direction. Hence in all cases we started with
a spiral/scroll wave near the centre of the domain, and variation of
the relative position of the initial spiral/scroll and a geometry
feature was done by variation of the feature. Placing the
  scroll wave, formed effectively in an infinite medium, into new
  geometic constraints did of course cause some initial fast transient
  while establishing its transmural structure before proceeding with
  the slow drift in the long dimensions. This transient appeared as a
  slight deviation of the tip trajectory compared to what it would be
  expected otherwise, during one or two initial rotation periods. This
  small deviation is not noticeable on the figures at the resolution
  we use in this paper, so we did not find it useful to eliminate or
  isolate this transient from the output data.

\end{widetext}

\end{document}